\documentstyle[emulateapj,psfig]{article}
\input{psfig}
\begin{document}
\title{Finding Black Holes with Microlensing}

\author{
Eric Agol\altaffilmark{1},
Marc Kamionkowski, L\'eon V. E. Koopmans, and Roger D. Blandford }
\affil{
California Institute of Technology, Mail Code 130-33, Pasadena, CA 91125 USA}
\altaffiltext{1}{Chandra Fellow, Email: agol@tapir.caltech.edu}

\begin{abstract}
The MACHO and OGLE collaborations have argued that the three longest-duration
bulge microlensing events are likely caused by nearby black holes, given
the small velocities measured with microlensing parallax and non-detection
of the lenses.  However, these events may be due to lensing by more numerous
lower-mass stars at greater distances.  We find a-posteriori probabilities 
of 76\%, 16\%, and 4\% that the three longest events are black holes,
assuming a Salpeter IMF and 40 $M_\odot$ cutoff for 
neutron-star-progenitors; the numbers depend strongly on
the assumed mass function, but favor a black hole for the longest event for
most standard IMFs.   The longest events ($> 600 $ days) 
have an a-priori $\sim$26\% probability to be black holes for a standard 
mass function.  We propose a new technique for measuring the 
lens mass function using the mass distribution of long events measured with 
ACS, VLTI, SIM, or GAIA.
\end{abstract}


\section{Introduction}

Counting the number of black-hole stellar-remnants in our Galaxy is 
complicated by their faintness.  Only a few dozen of an estimated
$10^7$--$10^9$ black-holes are observed as X-ray binaries (van den Heuvel 1992).
Black-hole numbers estimated from stellar-population synthesis are
inaccurate due to variations in IMF with metallicity, uncertainty
in the star-formation history, and uncertainty in the cutoff mass for
neutron-star progenitors.  To count the bulk of the black holes
(isolated or in wide binaries) requires a different observable 
signature, the best candidate being gravitational microlensing since 
it only relies on gravity.

Microlensing events are primarily characterized by 
the event timescale, $\hat t = 2 R_E/v$, where $R_E$ is the 
Einstein radius in the lens plane, $R_E=[4GMc^{-2}D_Sxy]^{1/2}$, 
$v$ is the velocity of the lens perpendicular to the 
observer-source axis, $M$ is the lensing mass,
$D_{L,S}$ are the distances to the lens or source, $x=D_L/D_S$ 
is the fraction of distance of the lens to the source, and $y=1-x$.  
The motion of the Earth causes magnification
fluctuations during long events, an
effect called ``microlensing parallax'' ($\mu$--$\pi$), allowing
one to measure the reduced velocity,
${\bf \hat v}=({\bf v_L}-x{\bf v_S})/y-{\bf v_\odot}$ (Gould 1992), 
where ${L,S,\odot}$ 
stand for the lens, source, and Sun and the velocities are perpendicular to 
the lensing axis.  Given $D_S$ and ${\bf \hat v}$, the lens mass is then 
solely a function of $x$,
\begin{equation}
M(x)=(\hat v\hat tc)^2y/(16GxD_S).
\end{equation}
The mass is $0$ for $x=1$, rising to $\infty$ for $x=0$, thus to estimate
$M$ requires knowing $x$.

Bennett et al. (2001, hereafter B01) and Mao et al. (2002, hereafter
M02) claim that the three
longest-duration microlensing events discovered toward the
bulge with $\mu-\pi$ measurements are likely black holes.  To
arrive at this conclusion, they assume that each lens is a
member of a population that has a velocity and spatial distribution
characteristic of the disk and bulge.  Here we in addition
assume that the lenses are drawn from a population that 
has a {\it mass function} characteristic of stellar and stellar
remnant populations in the disk and bulge, and is independent of $x$.  
This latter assumption is not true for young bright disk stars (which
do not contain much of the disk mass), but may be true for lower mass stars
and compact remnants in the disk and bulge.  We then find that the 
probability that these events are black holes is somewhat reduced.

In \S 2, we apply the analysis of B01 to MACHO-99-BLG-22
and estimate the implied total
black hole number.  In \S 3, we include a mass function in our
prior for the lens mass.  In
\S 4, we consider the timescale distribution for bulge
events and describe how mass measurements of long events
may provide an estimate of the lens mass function.  %We argue that
%the rate of occurrence of the long-duration microlensing events
%may be an anomaly that cannot be explained by ordinary stellar
%populations nor by the inclusion of a new high-mass population.
%We conclude by providing avenues toward possible resolution of
%these issues.

\section{Mass Estimate from Reduced Velocity Only}

B01 and M02 present seven microlensing 
events toward the bulge of the Galaxy with $\mu$--$\pi$
measurements as shown in Figure 1.   These events have greatly decreased 
$\chi^2$ when the $\mu$--$\pi$ effect is included and are
unlikely to be caused by a binary lens or source.
%We assume the rotation curve of the disk is flat, $v_L \sim v_\odot\sim
%220$ km/s.
%If one assumes that the source star in the bulge is motionless,
%${\bf v_S}\sim$ 0, then one can solve directly for the distance: 
%$x={\hat v/ (\hat v + v_\odot)}$.  All of 
%the Figure 1 events have $\hat v < 100$ km/s, indicating $x < 0.3$,
%placing a rough lower limit on each lens mass (Alcock et al. 1995).
%Disk stars have a velocity dispersion of $\sim$ 45 km/s
%and bulge stars $\sim$ 100 km/s, so B01
%have used a more careful version of the
%foregoing argument including lens and source velocity dispersions,
%the mass density distribution, and a constant prior for the lens
%mass to compute the probability of a lens lying at a given 
%distance (Alcock et al. 1995).  
B01 have computed a distance probability for MACHO-96-BLG-5 and 
MACHO-98-BLG-6 using a likelihood function which depends on the
observed reduced velocity and assumes the source is in the bulge.
Using equation (1) they find that the most likely 
masses for the their two longest events are $>3 M_\odot$ at 68\%
confidence.
Given that these lenses are not detected as main-sequence stars, they
argue that they must be black holes.  Using the likelihood
function of B01, we find that the mass probability for
MACHO-99-BLG-22 (OGLE-1999-BUL-32)
has $M=11_{-6}^{+12} M_\odot$ and a 81\% probability of being a black hole
using the best-fit parameters of Bennett et al. (2002, B02).  This differs
from the results of B02 due to different assumed disk and bulge
velocity dispersions.

%We have recomputed the maximum likelihood, ${\cal L}(M|{\bf \hat v})$, 
%for the masses of these three long-timescale events, assuming 
%the lenses have the spatial and velocity
%distributions of ordinary stellar populations.  
%${\cal L}(M|{\bf \hat v})$ we find that
%MACHO-96-BLG-5 has $M=4.7_{-2.3}^{+4.0} M_\odot$ (68\% confidence limits)
%with a 72\% probability of being a black hole (i.e. $M > 3 M_\odot$),
%while MACHO-98-BLG-6 has $M=4.6_{-2.0}^{+2.3} M_\odot$
%with a 75\% probability of being a black
%hole, and MACHO-99-BLG-22 (OGLE-1999-BUL-32)
%has $M=29_{-16}^{+34} M_\odot$ and a 97\% probability of being a black hole
%using the ${\bf \hat v}$ of Bennett et al. (2002).
%These values are slightly lower than B01 due to different model assumptions.

If all three events are due to disk black-holes, we can estimate the 
total number.
The detection efficiency, $\epsilon$, for events from 100--400 days is 
20\% (Alcock et al. 2000), which we assume also holds at longer $\hat t$
(these events are drawn from the alert sample which have a different
selection criterion, so a different $\epsilon$
may apply).  We have computed the timescale probability
for black-hole lensing events, and find that only $\sim$ 40\% have 
$\hat t > 1$ year, so $\epsilon \sim$ 13\%.
These three events yield $\tau\sim 5\times 10^{-7}(\epsilon/0.1)^{-1}$,
implying $N_{\rm total} \sim 5\times 10^8 (\langle M\rangle/9M_\odot)^{-1}
(\epsilon/0.1)^{-1}$ 
disk black holes.  This estimate is larger than estimates based upon the 
expected ratio of black-hole to neutron-star remnants, 
$\sim 10^8$, (Shapiro \& 
Teukolsky 1983, van den Heuvel 1992) and 
chemical enrichment by supernovae within the Milky Way, $\sim 2\times 10^8$ 
(Samland 1998).  This discrepancy indicates that either $\epsilon=
20$\% is too low, the IMF was much more top-heavy in the past,  or 
the events are due to more distant, low-mass stars.  
We next explore the third option.

\centerline{\psfig{file=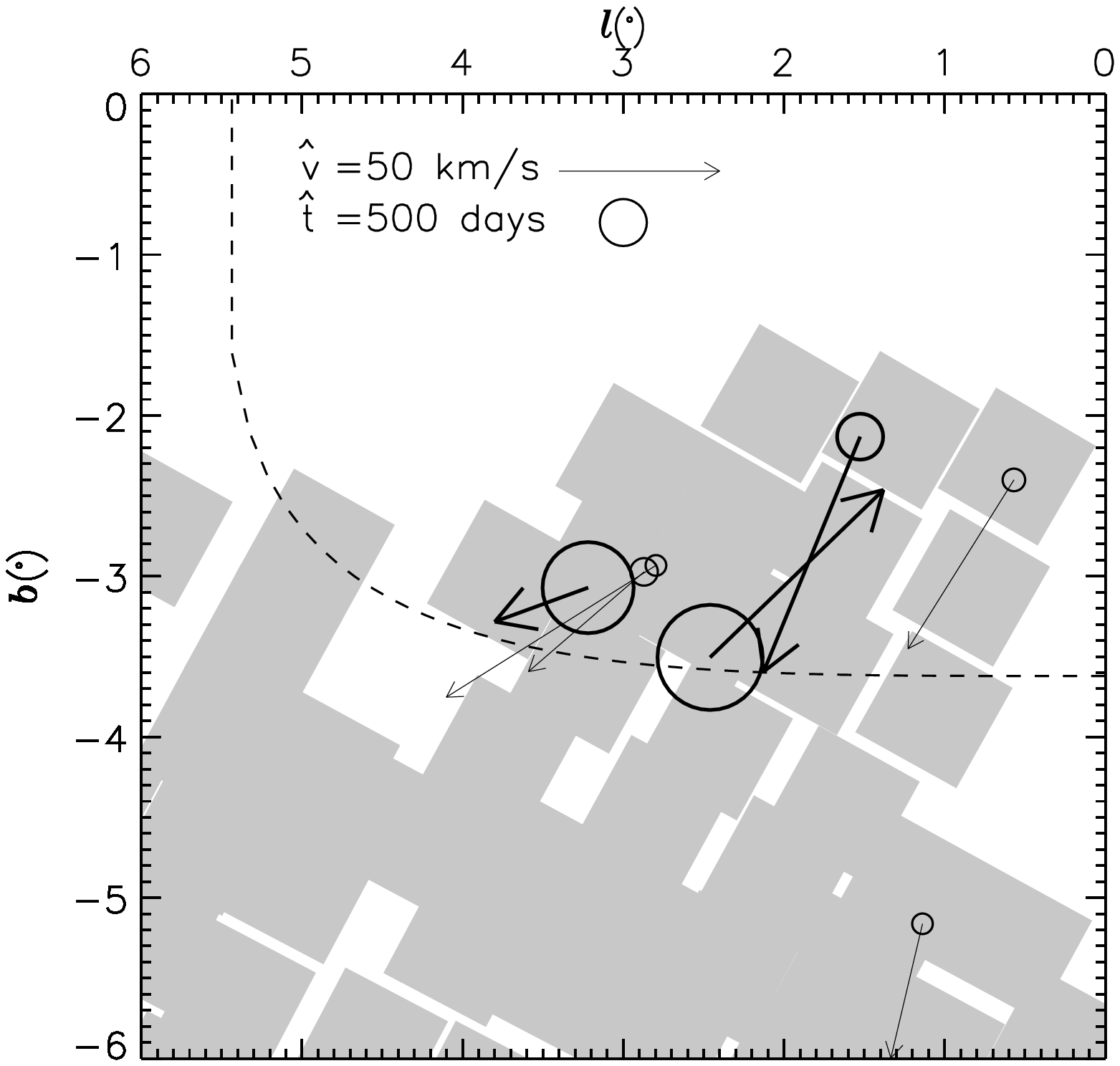,width=\hsize}} %FIGURE 1
\begin{normalsize}
Fig. 1 - The velocity vectors, ${\bf \hat v}$,
and event durations, $\hat t$ (proportional to circle size), for
seven $\mu$--$\pi$ microlensing
events from B01 and M02.
The shaded regions show the MACHO bulge fields, while the dashed
line shows the region where the bulge density is half of the
central bulge density in the sky plane at the distance of the bulge.
Boldface indicates three black hole candidate events. %FIGURE 1
\end{normalsize}
\vskip 3mm

\section{Mass Function Prior}

To recap, the small reduced velocities may indicate that the
three longest events most likely lie within 3 kpc and thus should have 
masses $> 3 M_\odot$.  However, if a lens is more distant, it is less
massive (equation 1) and thus drawn from a more abundant population,
which can compensate for the small probability of distant
lenses to have a small reduced velocity.  
Here we investigate the mass likelihood given ${\bf \hat v}$ {\it and}
$\hat t$ using the spatial probability $dn_L/dM=\rho(x)\phi(M)/\langle 
M\rangle$, where
$\phi(M)$ is the mass function of a 10 Gyr-old stellar population 
(independent of $x$), $\int \phi(M) dM =1$, and 
$\langle M\rangle =\int dM M \phi(M)$ is the mean stellar mass.

To compute $\phi(M)$, we start with the broken power-law IMF 
from Kroupa (2002), and, following Gould (2000),
convert stars with masses $M_{cms}<M<M_{cwd}$ into white dwarfs,
$M_{cwd}<M<M_{cns}$ into neutron stars, and $M_{cns} < M$ into black holes.
We set $M_{cms}=1M_\odot$ and $M_{cwd}=11M_\odot$ (Samland 1998), while
we vary $M_{cns}$.
We use Gaussian distributions for the white-dwarf and neutron-star mass
functions with $M_{wd}=0.5\pm0.1 M_\odot$ (Bragaglia et al. 1995) and 
$M_{ns}=1.35 \pm 0.04 M_\odot$ (Thorsett \& Chakrabarty 1999).
Black holes we describe with $\phi_{BH}\propto M^{-0.5}$ for $3 M_\odot
< M < 15 M_\odot$, consistent with the measured mass function of black
holes in X-ray binaries (which may have a different mass function than
isolated black holes).  We compute the relative numbers of each
compact remnant by varying the high-mass IMF slope $2<\beta<3$, where
$\phi_{IMF} \propto M^{-\beta}$, equating the number of each compact remnant
with the number of stars in the IMF with appropriate mass range.
We assume that this mass function applies
in both the disk and bulge.  

The number of black holes strongly depends on the slope of the 
IMF, $\beta$, and the maximum neutron-star-progenitor mass, $M_{cns}$.  
We choose as a fiducial value the Salpeter $\beta=2.35$,
consistent with the observed IMF (Kroupa 2002), and $M_{cns}=40 M_\odot$,
consistent with chemical-evolution models of the Galaxy (Samland 1998).
The resulting mass function has mass fractions
of 7\% in brown dwarfs, 77\% in main-sequence stars, 13\% in white dwarfs,
1\% in neutron stars, and 1.5\% in black holes.  This corresponds to
a total of about $10^9$ neutron stars and $2\times 10^8$ black holes in the 
Milky Way.  %We then use Bayes' theorem to provide an 
%estimate of the lens mass
%$p(M\mid \hat t,{\bf \hat v}) = {p(M) p(\hat t,{\bf \hat v}\mid M)/ 
%p(\hat t,{\bf \hat v})}$,
%assuming $p(M)= p(\hat t,{\bf \hat v})=1$, and normalizing the total
%probability to unity.  For a given $M$ and $D_S$, the lens 
%distance $x(M)$ is determined by equation (1), so we compute the 
%phase-space density at the observed 
%${\bf \hat v}$ and $\hat t$ and at the distance $x(M)$.  Multiplying
%by the mass function, $\phi$, source distance distribution, and event 
%cross section gives the probability as a function of mass.  

Using Bayes' theorem, the likelihood of a lens to be at a given 
distance, $x$, for fixed source distance $D_S$ is 
${\cal L}(x|\hat t, {\bf \hat v}) = {\cal L}(\hat t,{\bf \hat v}|x)$
for no prior on the distance and no errors on the
timescale and velocity measurements.
For a survey of duration $T$ covering a solid angle of $\Omega$, then
the solid angle cross section for magnification by $>30\%$ by a lens 
with reduced velocity $\hat v$ is $2 R_E \hat v y T/D_L^2$.  We must 
then multiply by the total number of lenses at that distance and 
integrate over the phase space of the velocity distribution of lenses 
and sources.  We include
lenses and sources in the disk and bulge using the bar model
described in Dwek et al. (1995) and Han \& Gould (1995) and a
double-exponential disk with scale length 3 kpc, scale height
325 pc, Galactic center distance 8 kpc, solar circular velocity
200 km/s, velocity ellipsoid $\sigma_z:\sigma_\phi:\sigma_r=1:1.3:2$,
$\sigma_z=17$ km/s at the solar circle with a scale length
of 6 kpc, and asymmetric drift of $\sigma_r^2/120$ km/s (Buchalter,
Kamionkowski \& Rich 1997).  The likelihood function becomes
\begin{eqnarray}
{\cal L}(\hat t,{\bf \hat v}|x)= N_S \int dM d{\bf v_S} d{\bf v_L}
{2 R_E \hat v y T } {dn_L \over dM} \cr
\times  D_S   f({\bf v_L}) f({\bf v_S})
\delta^2({\bf \hat v}({\bf v_S},{\bf v_L},x)-{\bf \hat v})
\delta(\hat t(M,\hat v,x)-\hat t).
\end{eqnarray}
%where $(dn_L/dM)dM$ is the number density of lenses between $M$ and $M+dM$.
The first delta function picks out the particular ${\bf \hat v}$
while the second delta function picks out the particular $\hat t$.
We convert the velocity delta function in ${\bf\hat v}$ to a
delta function in ${\bf v_L}$, then integrate over ${\bf v_L}$
and ${\bf v_S}$, and convert the second delta function to a delta
function in mass:  %We assume that the mass function is independent
%of position, and thus 
%\begin{equation}
%{dn_L \over dM} = {\phi(M)  \rho_L \over \langle M\rangle},
%\end{equation}
%where $\phi(M)dM$ is the fraction of lenses with mass between $M$ and
%$M+dM$ ($\int \phi(M) dM =1$) and $\langle M\rangle=\int M \phi(M) dM$ is 
%the average lens mass.  The probability becomes
\begin{eqnarray}
{\cal L}(\hat t,{\bf \hat v}|x) = {N_S 4 R_E  D_S y^3 T \hat v \over \hat t}
{M \phi(M) \over \langle M \rangle} \rho_L
{ e^{-\left[{v_l^2\over 2\sigma_l^2}+{v_b^2\over 2\sigma_b^2}\right]}
\over 2 \pi \sigma_l\sigma_b},
\end{eqnarray}
where ${\bf v}=(v_l,v_b)={\bf \bar v_L}-x{\bf \bar v_S}-y({\bf v_\odot} +{\bf \hat v})$,
$\sigma_{l,b}^2=\sigma_{L,l,b}^2+x^2\sigma_{S,l,b}^2$,
and $l,b$ denote the components directed in galactic longitude and 
latitude, respectively.  Comparing to the expression in B01, we see that the 
likelihood distributions agree if $\phi(M)\propto M^{-{3\over 2}}$.
We convert from the likelihood of a lens lying at a distance
$x$ to a likelihood in $M$ using equation (1).
We then integrate over sources at distance $D_S$ using a model for the
bulge and disk with luminosity functions measured with HST.
The final constraint is to require that the main-sequence lens stars not 
exceed the flux limits of B01 and M02 (we ignore the contribution
from red-giant lenses).

%To convert this into a probability for mass, one can write
%\begin{equation}
%p(M|\hat t,{\bf \hat v}) = p(\hat t,{\bf v}|M) = p(\hat t,{\bf \hat v}|x) |{dx \over dM}|.
%\end{equation}
%Since $x=M_0/(M+M_0)$ where $M_0=(\hat v \hat t c)^2/(16 G M_\odot D_S)$, where
%$M$ is measured in units of $M_\odot$, we find
%\begin{eqnarray}
%p(M|\hat t,{\bf \hat v}) = N_S 2 R_E (M_0\hat t)^{-1} D_S x^2y^3 T \hat v 
%\left[{M \phi(M) \over \langle M \rangle}\right] \cr \times \rho_L(x)
%(\pi \sigma_l \sigma_b)^{-1} e^{-\left[{v_l^2\over 2\sigma_l^2}+{v_b^2\over 2\sigma_b^2}\right]},
%\end{eqnarray}
%where $x=M_0/(M+M_0)$ should be substituted into this expression.

Figure 2 shows the computed mass probability for the above mass function.
The white-dwarf, neutron-star, main-sequence cutoff,
and black-hole mass cutoff show up as peaks in the probability 
distribution.  The greater number of lenses at small mass compensates 
for the decreased probability for low $\hat v$ events, biasing
the probability toward smaller mass.  The probability that each lens 
is a black hole, defined as $M > 3M_\odot$, is changed
from above:  4\% for MACHO-98-BLG-6, 16\% for MACHO-96-BLG-5, 
and 76\% for MACHO-99-BUL-22.  MACHO-98-BLG-6 and MACHO-96-BLG-5  are most 
likely  main-sequence or white-dwarf stars at $> 4$ kpc;  better flux limits
may rule out a main-sequence star.
These probabilities change by one order of magnitude depending on 
$\beta$ and $M_{cns}$, as shown in Table 1 (the labels are 
MACHO-XX-BLG-XX, while the first event is also OGLE-1999-BUL-32).
The first two columns describe the assumed mass function,
columns 3--5 give the black-hole probability for each event labeled
by their MACHO event number, while the last two columns give
the average of $M^{1/2}$ and $M^2$ for black holes divided by
the average for all stars given the assumed IMF.
%If black holes receive impulses at birth, 
%their number density and lensing timescale will be reduced, further reducing the
%black-hole probability.  Given these uncertainties, one cannot confidently 
%infer a lens mass $>3\,M_\odot$ based solely on a $\mu$--$\pi$ measurement.
\begin{figure*}
\centerline{\psfig{file=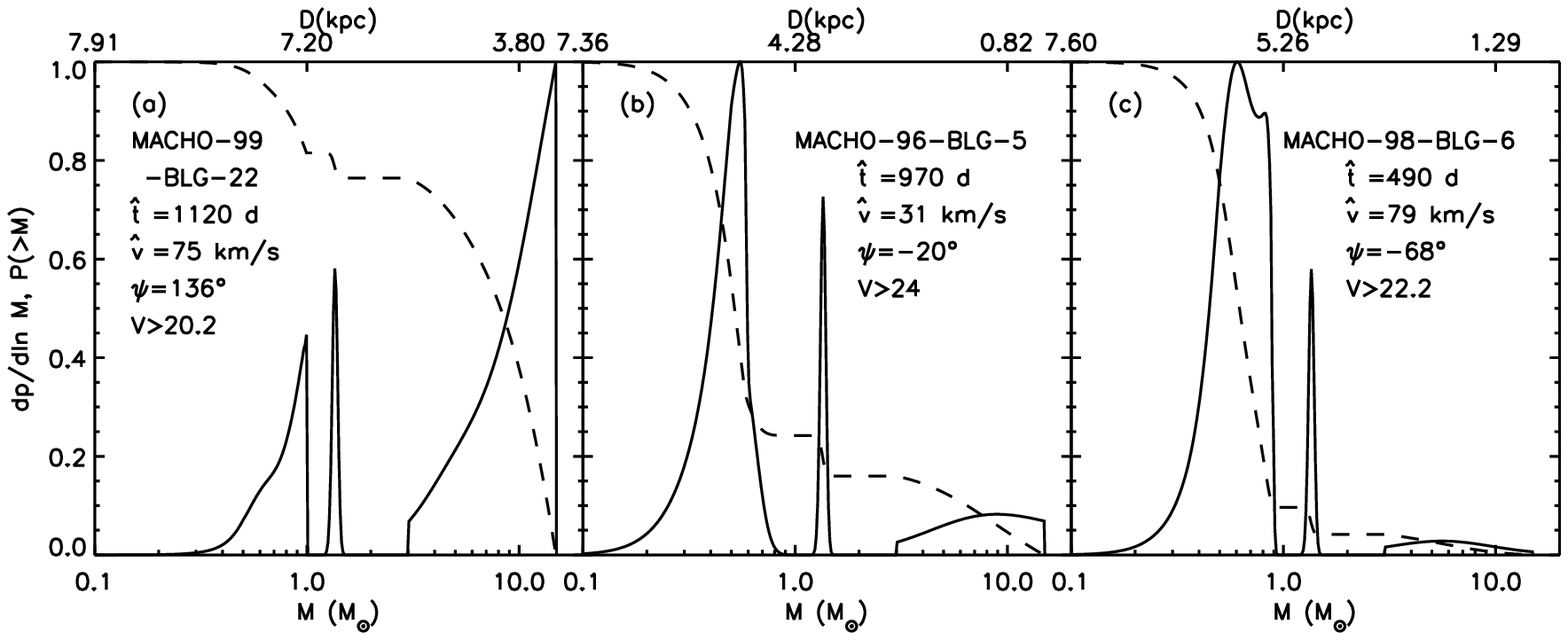,width=\hsize}} %FIGURE 2
%\centerline{\psfig{file=f2.eps,width=5in}} %FIGURE 2
\begin{normalsize}
Fig. 2 - The differential (solid) and integrated (dashed) likelihood of $M$ given
the observed ${\bf \hat v}$ and $\hat t$.
(a) MACHO-99-BLG-22 (OGLE-1999-BUL-32) (b) MACHO-96-BLG-5 (c) MACHO-98-BLG-6.
The mass function parameters are $\beta=2.35$ and $M_{cns}=40 M_\odot$.
The distance scale at the top of the plots shows $D_L(M)$ {\it if} $D_S=8$ kpc.
The angle $\psi$ is the velocity direction in Galactic coordinates measured from
the rotation direction towards the north Galactic pole.
The $V$-band flux limits are from (a) M02 (b) Bennett et al. (2002) 
and (c) B01. %FIGURE 2
\end{normalsize}
\end{figure*}

\section{Probability of Long-Timescale Lensing}

%Although we have now shown that the long-duration $\mu$--$\pi$
%events can be consistent with lower-mass lenses at larger
%distances, there is still a problem with the timescale
%distribution.  As we will show, this is difficult to explain
%with either a conventional stellar population or with a new
%population of black holes.  
%
%For a fixed mass, long-timescale events are produced by lenses 
%with small sky velocity;  however, the smaller the velocity, 
%the smaller the volume of phase space.   Consider events
%with sky velocities in the range $v$ to $v+dv$.  Lenses cover an
%area of the sky per unit time given by $2R_Ev$ (for
%$R_E \ll v T$), while
%$v=2R_E/\hat t$, $dv \propto d\hat t / \hat t^2$, so the 
%differential probability of lensing at long timescale scales as 
%$(2R_Ev) v dv/d\hat t\propto \hat t^{-4}$ (Mao \& Paczy\'nski 
%1996).  
We next compute the expected distribution of events as a function of
timescale.
For long timescales the a-priori differential probability distribution
scales as $\hat t^{-4}$ (Mao \& Paczynski 1996).
For solar-mass stars in the disk and bulge this behavior occurs for
$\hat t > 200$ days.
We can rescale the probability distribution for different masses
\begin{equation}
{d^2p\over dM d\hat t} =\langle M\rangle^{-1}{\phi(M)}{dp(M_\odot)\over d\left(\hat t M^{-1/2}\right)},
\end{equation}
where $M$ is measured in units of $M_\odot$ and
${dp(M_\odot)/d\hat t}$ is the timescale probability
distribution assuming all lenses have mass $M_\odot$.  Since at
long timescales ${dp(M_\odot)/ d\hat t}$ scales as $\hat t^{-4}$,
we can rescale this equation and integrate over mass
\begin{equation}
{dp\over d\hat t}= \int_{M_1}^{M_2}{d^2p\over dM d\hat t} dM
\propto \hat t^{-4}\int_{M_1}^{M_2}dM{\phi(M)} M^2,
\end{equation}
for $\hat t > M_2^{1/2}$ 200 days.
The fraction of all microlensing events as a function of mass
scales as $M^{1/2}\phi(M)$, while the fraction of events in the
long-timescale tail scales as $M^2\phi(M)$, strongly favoring
black holes.  Table 1 shows the average of $M^{1/2}$ and $M^2$
for black holes divided by mass-function average,
showing that the fraction of black-hole events in the
long-timescale tail is increased by one to two orders of
magnitude as compared to the total number of events.  {\it Thus,
with a measurement of the mass distribution of events at long
timescales, one can directly infer the mass function if it is
independent of the location in phase space.}
\begin{center}
\centerline{\bf Table 1: Black Hole Probabilities}
\begin{tabular}{|c|c|ccc|c|c|}
\hline
$\beta$ & $M_{cns}$ &&P(\%)& & ${\langle
M_{BH}^{1/2}\rangle\over\langle M^{1/2}\rangle}$ & ${\langle M_{BH}^2\rangle\over
\langle M^2\rangle}$  \\
\hline
 &  & 99-22& 96-5 & 98-6 & &\\
\hline
\hline
2    & 20 & 97   & 65 & 30 &  3E-2 & 0.78 \\
2    & 40 & 92   & 45 & 17 &  1.5E-2 & 0.64 \\
2.35 & 20 & 90   & 33 & 10 &  7E-3& 0.48\\
2.35 & 40 & 76   & 16 & 4  &  3E-3& 0.26\\
3    & 20 & 46   & 4  & 1  &  5E-4& 0.07\\
3    & 40 & 18   & 1  &0.2 &  1E-4& 0.02\\
\hline
\end{tabular}
\end{center}

We have computed the probability distribution for lensing by
solar-mass stars (see Han \& Gould 1995, 1996 for the computation 
technique), and then convolved with our assumed mass function to
give the timescale distribution. This is compared to the
observed distribution of MACHO events in Figure 3.
We have binned the 252 alert events from MACHO 
(http://darkstar.astro.washington.edu) and then computed 
$\hat t^2 dn/d\hat t$ as $N_i \hat t_i^2 /(\epsilon_i \Delta\hat t_i)$, 
where $i$ labels the number of the bin, errors are Poisson.  
Since the efficiencies, $\epsilon_i$, have not been measured
for the alert events, we have used the efficiency curve
from Alcock et al. (2000), extrapolating to longer timescale.
We note that the alert events timescales may be affected by blending and
that a full analysis of efficiencies is required.
The rise at long timescale may indicate that 20\% underestimates
the efficiency or that the longest two events are a statistical
fluke.  An efficiency of 100\% is indicated for the the last three bins 
by an arrow.  Decreasing $\beta$ to 2 improves agreement for the shorter
timescales, but the longest event still lies well above the predicted 
value, so it may be an improbable event.  

We have tried to explain the rise at long timescale
by varying the assumptions in the Galactic kinematics and
density, but all reasonable modifications fail.  To explain this 
discrepancy with a different mass function requires stars of mass 
$\sim 100 M_\odot$ since the timescale scales as $M^{1/2}$
and the probability distribution peaks at $\sim 100$ days for
$1 M_\odot$ lenses.  However, $100 M_\odot$ lenses are ruled 
out by the $\mu$--$\pi$ observations which indicate that the
three longest timescale events would have to be within a few
hundred parsecs.  This would require an even larger black hole mass
density to explain the number of these events, $\sim 1 M_\odot$ pc$^{-3}$,
which exceeds the Oort limit.  Also, for events near the
peak of the timescale distribution for $100 M_\odot$ lenses, 
$\hat v$ should be much larger than observed.
\centerline{\psfig{file=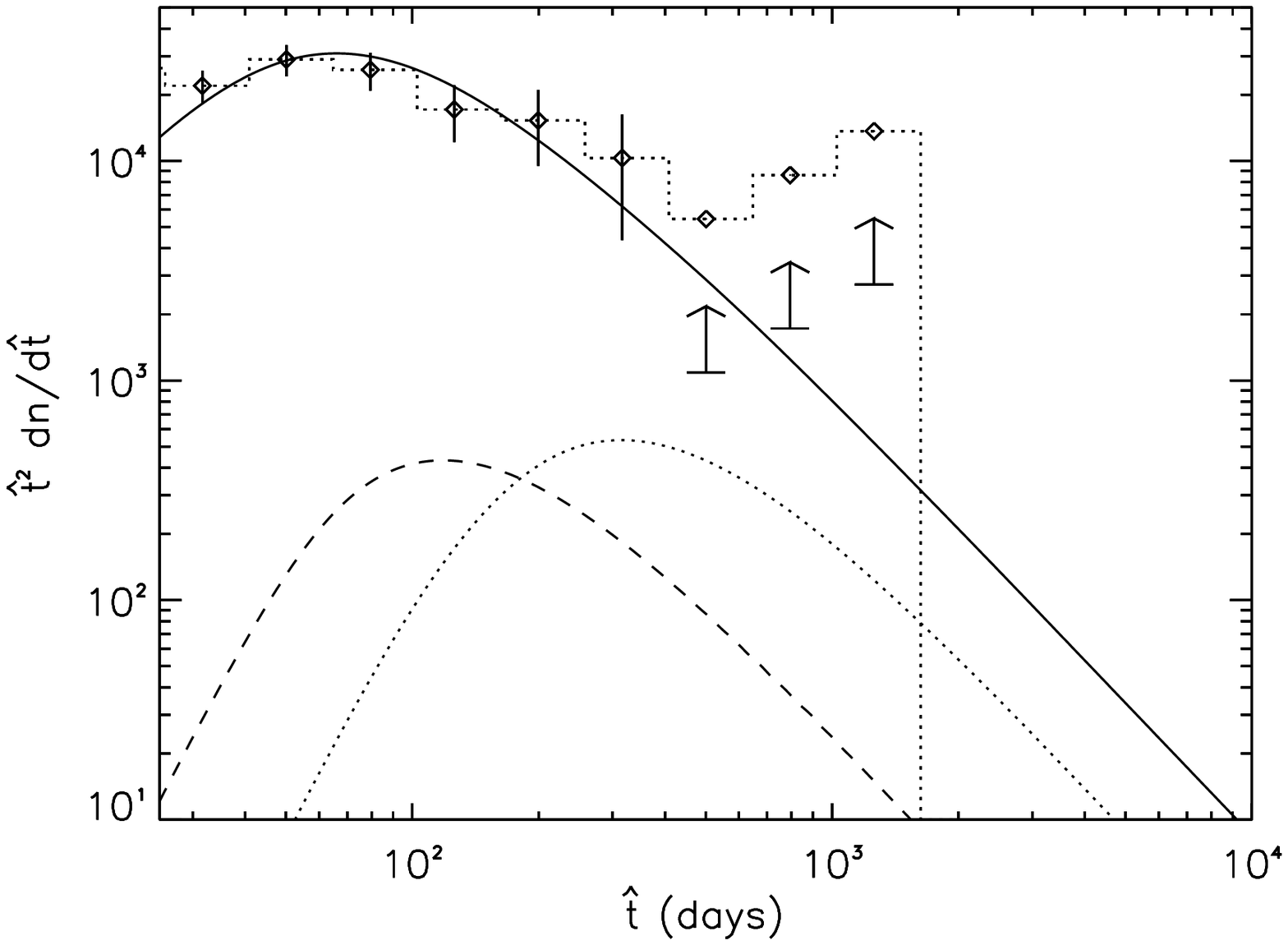,width=\hsize}} %FIGURE 3
\begin{normalsize}
%Fig. 3 - The distribution of event timescales,
%$\hat t^2 dn/d\hat t$.  The histogram shows the MACHO alert events,
%while the lines show the predicted distribution for all masses (solid),
%neutron stars (dashed), and black holes (dotted) for the fiducial IMF.%FIGURE 3
Fig. 3 - The distribution of event timescales,
$\hat t^2 dn/d\hat t$.  The histogram shows the MACHO alert events,
while the lines show the predicted distribution for all masses (solid),
neutron stars (dashed), and black holes (dotted) for the fiducial IMF. %FIGURE 3
\end{normalsize}
\vskip 3mm

%We have used regularized inversion of equation (2) to solve for a mass
%function given the observed timescale distribution, finding that the
%data can be fit as a single $\beta=2$ power-law from $M_1=0.1 M_\odot$ to
%$M_2=100 M_\odot$.  Using this mass function as a prior, we find a black-hole
%probability which is similar to the $\beta=2.35$, $M_{cns}=20 M_\odot$
%case.

\section{Conclusions}

%We have shown that the longest microlensing events are likely
%caused by stars with $M<3M_\odot$ for a range of IMFs. 
We have shown that including the timescale as a constraint on the
black hole probability may change the conclusion as to whether
or not the lens is a black hole.  The probability is most robust for
MACHO-99-BLG-22, which is most likely a black hole for the
range of mass functions we have explored, while the probability
for MACHO-98-BLG-6 and MACHO-96-BLG-5 is strongly dependent
on the assumed mass function.
%For a given $\mu$--$\pi$ event, the lens is less massive if it is
%more distant, and such lenses are more numerous.  This 
%compensates for the small probability of a lens and source having
%the appropriate velocities to cause the small observed velocities.
%We find that the black-hole hypothesis is favored at only the 8\% and 3\%
%levels for the two MACHO events, but is more strongly favored at 
%20\% confidence for the longest
%event discovered by OGLE.  However,
%this conclusion depends strongly on the parameters of the mass function
%that we assume (Table 1).  
A $\mu$--$\pi$ measurement does not provide
sufficient information to estimate the mass of a given event, 
but may result in interesting limits on the mass.
%Indeed, of order $\langle M^{1/2}\rangle_{BH}/\langle M^{1/2}
%\rangle = 0.3$\% events, or $\sim 1$ of the MACHO or OGLE events,
%may have been a black hole if $\beta=2.35$ and 
%$M_{cns}=40 M_\odot$, but we cannot determine which one.

The contribution of a given population of stars to the long-timescale
tail ($\hat t > 200 M^{1/2}$ days) of the lensing distribution depends 
on $\langle M^2 \rangle$ for that population.
Due to their larger mass, $\sim$ 26\% of events with $\hat t > 600$ 
days should be black-hole lenses for $\beta=2.35$ and $M_{cns}=40M_\odot$;  
this fraction may be reduced if black holes have a velocity dispersion much 
larger than that of disk and bulge stars.  To constrain $\beta$ and
$M_{cns}$ with the timescale distribution will require more events and
a careful estimate of the microlens detection efficiency as a function
of $\hat t$.  The parallax events may have an increased efficiency since
they are less affected by observing gaps and have a wider cross section
due to the Earth's motion (B01).

The OGLE III project may detect $10^3$ microlensing events toward the
bulge each year (Paczy\'nski, priv. comm.), which should include $\sim 3$ 
black-hole lensing events
and $\sim 5$ neutron-star events for $\beta=2.35$ and $M_{cns}=40 M_\odot$.
Since the timescale of black-hole events is longer, $\mu$--$\pi$ measurements 
are feasible from the ground (as demonstrated by B01).
If these can be followed up with astrometric observations, then the mass
can be determined when combined with $\mu$--$\pi$ information (Paczy\'nski 1998,
Gould 2001, Boden, Shao, \& van Buren 1998).
Since the typical mass of black holes, $\sim 9 M_\odot$, is much higher
than the typical lens mass, $\sim 0.5 M_\odot$, the typical astrometric
signal for black-hole lens events with a source in the bulge will be 
$\sim$ 3 mas which will easily be measured with ACS (M02), VLTI
(by 2004, Delplancke et al. 2001), SIM (by 2009), or GAIA (by 2010-12, 
Belokurov \& Evans 2002),
compared to 0.7 mas for a typical main-sequence lens.  
The deviation of the centroid is $\delta {\bf \theta} =
\theta_E {\bf u}/(u^2+2)$,
where $u$ is the normalized impact parameter as a function of time
and $\theta_E$ is the Einstein angle (Paczy\'nski 1998).  Since $u$ is 
determined by the
photometric light curve, observations of $\delta {\bf \theta}$ at two different $u$
will suffice to determine $\theta_E$. 
Combined with $\hat R_E=\hat v \hat t/2$ measured photometrically from the 
$\mu$--$\pi$ will allow a determination of the mass of the lens,
$M=c^2\hat R_E \theta_E/(4G)$.
With mass measurements for many long-timescale events, the mass function 
can be deduced since the probability for lensing is 
proportional to $\hat t^{-4}M^2\phi(M)$ for a phase-space distribution that
is independent of mass (equation 3).  The $M^2$ dependence means that 
detection of black holes will be favored so that one can constrain 
$\beta$ and $M_{cns}$.

Finally, direct observations of the lenses provide another
avenue toward determining their nature.  If they
are black holes, they may accrete from the interstellar
medium or from a companion wind.  With plausible assumptions,
the X-ray radiation could be detectable with current space-based
observatories (Agol \& Kamionkowski 2001).

\section*{ACKNOWLEDGMENTS}

We thank Dave Bennett and Shude Mao for constructive comments. This 
work was supported in part by NSF AAST-0096023, NASA NAG5-8506, and DoE 
DE-FG03-92-ER40701.  EA was supported by NASA through Chandra
Postdoctoral Fellowship Award Number PF0-10013 issued by the Chandra
X-ray Observatory Center, which is operated by the Smithsonian 
Astrophysical Observatory for and on behalf of NASA under contract NAS8-39073.

%\figcaption{The velocity vectors, ${\bf \hat v}$,
%and event durations, $\hat t$ (proportional to circle size), for
%seven $\mu$--$\pi$ microlensing
%events from B01 and M02.
%The shaded regions show the MACHO bulge fields, while the dashed
%line shows the region where the bulge density is half of the
%central bulge density in the sky plane at the distance of the bulge.
%Boldface indicates three black hole candidate events.} %FIGURE 1
%
%
 
%\begin{figure}
%\plotone{f1.eps} %FIGURE 1
%\end{figure}

%\begin{figure}
%\plotone{f2.eps} %FIGURE 2
%\end{figure}

%\begin{figure}
%\plotone{f3.eps} %FIGURE 3
%\end{figure}

\end{document}